\documentclass[aps,showpacs,superscriptaddress,preprint]{revtex4}%
\usepackage{mathrsfs}
\usepackage{amsfonts}
\usepackage{amsmath}
\usepackage{amssymb}
\usepackage{graphicx}
\usepackage{xcolor}%
\begin{document}
\title{Theory of multiple magnetic scattering for quasiparticles on a gapless
topological insulator surface}
\author{Zhen-Guo Fu}
\affiliation{SKLSM, Institute of Semiconductors, CAS, P. O. Box 912, Beijing 100083, China}
\affiliation{LCP, Institute of Applied Physics and Computational Mathematics, P.O. Box
8009, Beijing 100088, China}
\author{Ping Zhang}
\thanks{zhang\_ping@iapcm.ac.cn}
\affiliation{LCP, Institute of Applied Physics and Computational Mathematics, P.O. Box
8009, Beijing 100088, China}
\author{Zhigang Wang}
\affiliation{LCP, Institute of Applied Physics and Computational Mathematics, P.O. Box
8009, Beijing 100088, China}
\author{Fawei Zheng}
\affiliation{LCP, Institute of Applied Physics and Computational Mathematics, P.O. Box
8009, Beijing 100088, China}
\author{Shu-Shen Li}
\thanks{sslee@semi.ac.cn}
\affiliation{SKLSM, Institute of Semiconductors, CAS, P. O. Box 912, Beijing 100083, China}

\begin{abstract}
We develop a general low-energy multiple-scattering partial-wave theory for
gapless topological insulator (TI) surfaces in the presence of magnetic
impurities. As applications, we discuss the differential cross section (CS)
$d\Lambda/d\varphi$, the total CS $\Lambda_{tot}$, the Hall component of
resistivity $\Omega$, and inverse momentum relaxation time $\Gamma_{M}$ for
single- and two-centered magnetic scattering. We show that differing from the
nonmagnetic impurity scattering, $s\mathtt{-}$wave approximation is not
advisable and convergent in the present case. The symmetry of CS is reduced
and the backscattering occurs and becomes stronger with increasing the
effective magnetic moment $M$ of single magnetic impurity. We show a non-zero
perpendicular resistivity component $\Omega$, which may be useful for tuning
the Hall voltage of the sample. Consistent with the analysis of $d\Lambda
/d\varphi$, by comparing $\Gamma_{M}$ with $\Lambda_{tot}$, we can determine
different weights of backscattering and forward scattering. Similar to CS,
$\Omega$ and $\Gamma_{M}$ also exhibit oscillating behavior for multiple
magnetic scattering centers due to interference effect.

\end{abstract}

\maketitle

\section{Introduction}

A topic of fundamental importance in condensed matter physics is how the
presence of defects or impurities induce strong modifications on the local
electronic properties of crystalline solids. These modifications, with the
stunning development of scanning tunneling microscopy (STM), have been
extensively investigated on metal surfaces where they are well known as
Friedel oscillations and manifest as standing waves in the local electronic
density spanning regions up to $\sim$10 nm from the defects on the metal
surfaces. One kind of particularly suitable prototype that have been used in a
large amount of STM measurements to study the effects of impurity and the
formation of adsorbate superstructure are the (111) surfaces of noble metals,
on which the surface-state electrons form a two-dimensional (2D) nearly
free-electron gas. These Shockley-type surface states are dispersed as
$\epsilon=\hbar^{2}k^{2}/2m_{eff}$ (measured relative to the bottom of the
surface-state band) and localized in narrow band gaps in the center of the
first Brillouin zone of the (111)-projected bulk band structure. Thus they
have extremely small Fermi wave vectors $k_{f}=\sqrt{2m_{eff}\epsilon_{F}%
}/\hbar$ and consequently the Friedel oscillations of the surface state have a
significantly larger wavelength than those of the bulk states.

Recently, topological insulator (TI) has attracted tremendous experimental and
theoretical studies \cite{Hasan,Qi2011}. Unlike (111) surfaces of noble
metals, a peculiar characteristic of TI is the presence of strong spin-orbit
coupling (SOC), which results in a variety of unique properties. One
intriguing fact is that the ideal TI surface is described at low energies by a
2D massless-Dirac wave equation with an additional locking between momentum
and spin of surface electron. Because of the Dirac spectrum and SOC induced
fermionic chirality, the impurity scattering effect in TIs is naturally
expected to display novel behavior that should be absent from the conventional
semiconductor or metal-surface 2D electron gases. Many experimental and
theoretical efforts towards this issue have been payed. The anomalous Friedel
oscillations in the vicinity of a single localized impurity
\cite{Liu2009,Balatsky2010,Balatsky2011,Balatsky2012}, as well as the
identification of the nature and the precise location of impurities on TI
surface using STM \cite{Xue2009,Roushan2009,Xue2011,Alpichshev2011,Biswas2011}%
, have been discussed. However, when impurities are located close to each
other, multiple scattering effects should be important, such as the issue of
the long-range interactions between the adsorbates mediated by the Dirac
electrons of TI surface \cite{ZGFu2011,ZGFu20112}. In particular, since the
quasiparticle's spin is strongly coupled to its momentum, quantum interference
between different spin states during multiple scattering process could then
display prominent phenomena such as electric conductance weak
(anti-)localization \cite{He} and Aharonov-Bohm effect \cite{Fu2011} in STM signals.

In the presence of the time-reversal symmetry (TRS), the backscattering
induced by nonmagnetic impurities is forbidden on the gapless TI surface
because of a $\pi$ Berry phase associated with the $2\pi$ adiabatic rotation
of Dirac electron spin along the Fermi energy surface. However, considering
the magnetic impurities on the gapless TI surface, one would like to observe
the backscattering since the TRS is broken. Many efforts have been devoted to
exploring this issue. For example, very recently, quasiparticle interference
induced by a magnetic Co adatom on gapless Bi$_{2}$Se$_{3}$ surface has been
found in STM experiments \cite{Ye2011}. Furthermore, a magnetic field can be
generated when the TI sample is deposited on a lithographically patterned
ferromagnetic layer, which could also induce backscattering of massless Dirac
electrons in TI \cite{Zazunov2010}.

Because of its importance both from basic point of interest and to TI-based
chemical catalysis and electronics, in the present paper we address this issue
by presenting a first attempt at the theoretical evaluation of the multiple
scattering problem of the massless Dirac electrons on the TI surface in the
presence of localized and identical magnetic impurities. Specially, we present
the analytical expressions for multiple partial-wave scattering of massless
Dirac electrons with magnetic impurities, based on which the asymptotic
multiple scattering amplitude for random arraying magnetic impurities are
obtained under the particular large distance approximations (the identical
impurities are treated as a large scattering center). The differential and
total cross sections (CSs), the inverse momentum relaxation time, and the Hall
component of resistivity are discussed. We find that differing from the
nonmagnetic scattering, for the magnetic impurity scattering, the CS is not
convergent under $s\mathtt{-}$wave approximation. Therefore, higher partial
waves should be introduced. For the single magnetic impurity scattering, we
show the fact that the backscattering becomes much stronger when increasing
the effective magnetic moment $M$. By comparing the inverse momentum
relaxation time with total CS, we can determine whether there exist more
backscattering than forward scattering or not. Similar to CS, the inverse
momentum relaxation time and Hall factor display oscillating behavior for
multiple magnetic scattering centers due to interference.

\section{Model and theory}

The eigenstates of the effective low-energy Hamiltonian of TI surface near the
Dirac point \cite{Zhang2009}
\begin{equation}
H_{0}\mathtt{=}\hbar v_{f}\left(  \boldsymbol{\sigma}\mathtt{\times
}\boldsymbol{k}\right)  \mathtt{\cdot}\hat{z}%
\end{equation}
are given by the spinors $\Psi_{0,\pm}\left(  \boldsymbol{r}\right)
\mathtt{=}\frac{e^{i\boldsymbol{k}\cdot\boldsymbol{r}}}{\sqrt{2}}\left(
\begin{array}
[c]{cc}%
1, & \mp ie^{i\theta_{\boldsymbol{k}}}%
\end{array}
\right)  ^{\text{T}}$, where the upper/lower sign corresponds to the
electron/hole part of the spectrum. Here, $v_{f}\sim5\times10^{5}$ m/s is the
Fermi velocity, $\boldsymbol{\sigma}$ are Pauli matrices, and $\boldsymbol{k}$
is the in-plane wavevector. The impurity potential can be expressed as
\begin{equation}
V_{i}\mathtt{=}\frac{1}{2}J_{i}\boldsymbol{S}_{i}\mathtt{\cdot}%
\boldsymbol{\sigma}\Theta\left(  a-r\right)  ,
\end{equation}
where $\boldsymbol{S}_{i}\mathtt{=}S\boldsymbol{n}_{i}$ is the classical spin
(with its orientation vector $\boldsymbol{n}_{i}$) of the $i$th magnetic
impurity, $J_{i}$ is the exchange coupling strengths, $a$ is the radius of the
scatterer, and $\Theta\left(  r\right)  $ is the Heaviside function. If the
measurements are performed at a temperature higher than the Kondo temperature
\cite{Cui}, the coupling between impurity spins will not exceed the critical
value $J_{cr}$ before a Kondo effect occurs. In this work we assume that the
exchange coupling $J_{i}<J_{cr}$, so that the Ruderman-Kittel-Kasuya-Yosida
interactions between impurity spins and Kondo screening of the impurity spin
by the band electrons are neglected, and the impurity spin acts as a classical
local magnetic moment under mean-field approximation
\cite{Liu2009,Balatsky2010}.

In order to obtain the analytical expressions of wavefunctions, we just
consider the component of the classical spin along the normal line of TI
surface, i.e., $V_{i}\mathtt{=}M\sigma_{z}\Theta\left(  a-r\right)  $. The
fact that a magnetic Co impurity with only perpendicular spin component on the
TI surface does not open a gap has been experimentally observed \cite{Ye2011}.
To develop a scattering theory from localized, cylindrically-symmetric
scatterers, it is convenient to resolve the problem in cylindrical
coordinates. By considering the continuity of the wavefunction at the boundary
of the magnetic scattering potential, one can immediately obtain the
analytical expression of the scattered wave, written as
\begin{equation}
\Psi_{\text{sc}}\left(  \boldsymbol{r},\boldsymbol{k},\pm\right)
\mathtt{=}s_{0}G_{+0}T_{0}^{-}\Phi_{\pm}^{in}\mathtt{+}\sum_{l=1}^{\infty
}\left[  s_{l}G_{+l}T_{l}^{-}\mathtt{+}s_{-l}G_{-l}T_{l}^{+}\right]  \Phi
_{\pm}^{in}. \label{scw}%
\end{equation}
Here, $\Phi_{\pm}^{in}$ denotes the incident plane-wave centered about a
single scatterer located at $\boldsymbol{r}_{n}$. The cylindrically-symmetric
Green's functions take the form%
\begin{align}
G_{+l}  &  =\frac{i^{l}e^{il\theta_{n}}}{2}\left(
\begin{array}
[c]{cc}%
H_{l}^{(1)}\left(  k\rho_{n}\right)  & \pm H_{l}^{(1)}\left(  k\rho_{n}\right)
\\
\pm H_{l+1}^{(1)}\left(  k\rho_{n}\right)  e^{i\theta_{n}} & H_{l+1}%
^{(1)}\left(  k\rho_{n}\right)  e^{i\theta_{n}}%
\end{array}
\right)  ,\\
G_{-l}  &  =\frac{i^{l}e^{-il\theta_{n}}}{2}\left(
\begin{array}
[c]{cc}%
H_{l}^{(1)}\left(  k\rho_{n}\right)  & \mp H_{l}^{(1)}\left(  k\rho_{n}\right)
\\
\mp H_{l-1}^{(1)}\left(  k\rho_{n}\right)  e^{i\theta_{n}} & H_{l-1}%
^{(1)}\left(  k\rho_{n}\right)  e^{i\theta_{n}}%
\end{array}
\right)  ,
\end{align}
for $\left\vert \epsilon\right\vert >M$, where upper and lower signs in the
right side of these expressions denote the $\epsilon>0$ and $\epsilon<0$ parts
of the spectrum, $k=\frac{\epsilon}{\hbar v_{f}}$, $\boldsymbol{\rho}%
_{n}\mathtt{=}\boldsymbol{r}\mathtt{-}\boldsymbol{r}_{n}$ and $e^{i\theta_{n}%
}\mathtt{=}\frac{\boldsymbol{\rho}_{n}\cdot\left(  \hat{x}+i\hat{y}\right)
}{\rho_{n}}$. For the energy regin of $\left\vert \epsilon\right\vert <M$, the
Hankel functions $H_{l}^{(1)}\left(  k\rho_{n}\right)  $ in $G_{\pm l}$ should
be replaced by the modified Bessel functions of first kind $I_{l}\left(
k\rho_{n}\right)  $. The scattering amplitude is expressed as
\begin{equation}
s_{l}=\frac{A_{+}J_{l}\left(  k^{\prime}a\right)  J_{l+1}\left(  ka\right)
-A_{-}J_{l}\left(  ka\right)  J_{l+1}\left(  k^{\prime}a\right)  }{A_{-}%
H_{l}^{(1)}\left(  ka\right)  J_{l+1}\left(  k^{\prime}a\right)  -A_{+}%
H_{l+1}^{(1)}\left(  ka\right)  J_{l}\left(  k^{\prime}a\right)  } \label{ss}%
\end{equation}
for $\left\vert \epsilon\right\vert >M,$ where $A_{\pm}=\sqrt{\left\vert
\epsilon\pm M\right\vert }$, $k^{\prime}=\frac{\sqrt{\left\vert \epsilon
^{2}-M^{2}\right\vert }}{\hbar v_{f}}$, and $J_{l}$ is the Bessel function of
order $l$. Whereas $s_{l}$ should also be replaced by
\begin{equation}
\widetilde{s}_{l}=\frac{A_{+}I_{l}\left(  k^{\prime}a\right)  J_{l+1}\left(
ka\right)  +A_{-}J_{l}\left(  ka\right)  I_{l+1}\left(  k^{\prime}a\right)
}{-A_{-}H_{l}^{(1)}\left(  ka\right)  I_{l+1}\left(  k^{\prime}a\right)
-A_{+}H_{l+1}^{(1)}\left(  ka\right)  I_{l}\left(  k^{\prime}a\right)  }\text{
} \label{ss2}%
\end{equation}
for the case of for $\left\vert \epsilon\right\vert <M.$ Note that $s_{l}$
($\widetilde{s}_{l}$) satisfies the unitarity condition $\operatorname{Re}%
[s_{l}]=-\left\vert s_{l}\right\vert ^{2}$ ($\operatorname{Re}[\widetilde
{s}_{l}]=-\left\vert \widetilde{s}_{l}\right\vert ^{2}$), and
\begin{equation}
\lim_{\epsilon\rightarrow M^{+}}s_{l}=\lim_{\epsilon\rightarrow M^{-}%
}\widetilde{s}_{l}=-J_{l+1}\left(  ka\right)  /H_{l+1}\left(  ka\right)
\label{limit}%
\end{equation}
for all $l$. The $l^{th}$\texttt{-}partial-wave $t\mathtt{-}$matrix is
$T_{l}^{\pm}\mathtt{=}$diag$(%
\begin{array}
[c]{cc}%
\hat{P}_{l}^{\pm}, & \mp i\hat{P}_{l\mp1}^{\pm}%
\end{array}
)$ with $\hat{P}_{l}^{\pm}=\frac{e^{\pm il\theta}}{i^{l}k^{l}}(\partial_{r}%
\pm\frac{i}{r}\partial_{\theta})^{l}$. A detailed derivation is given in
Appendix A.

It is easy to extend scattering theory of massless Dirac fermions to the
realistic and reasonable case of multiple magnetic impurities, where the
quantum interference effect in the propagation process of Dirac fermions on TI
surface can be observed. This has not been discussed in previous studies, such
as Ref. \cite{Zazunov2010}. Taking into account all of partial waves, for $N$
magnetic scatterers located at positions $\boldsymbol{r}_{1},\boldsymbol{r}%
_{2},\cdots\boldsymbol{r}_{N}$, the scattered wavefunction is given by
\begin{equation}
\Psi_{\text{sc}}\left(  \boldsymbol{r}\right)  =\mathbb{G}\left(
\boldsymbol{r}\right)  \mathbf{S}D^{-1}\vec{\phi}. \label{10}%
\end{equation}
Here, $\mathbb{G}\left(  \boldsymbol{r}\right)  $ (a $2\mathtt{\times
}2N\left(  2l_{\max}\mathtt{+}1\right)  $ matrix) contains the propagation
information from detector to impurities $G_{\pm l}\left(  \boldsymbol{r}%
,\boldsymbol{r}_{i}\right)  $. $S$ is a diagonal matrix with nonzero element
$s_{\pm l}^{\left(  n\right)  }$. The $\mathbf{G}$ matrix, is constructed by
$T_{l}^{-}\left[  G_{\pm l^{\prime}}\left(  n,m\right)  \right]  $, describing
the propagation between impurities. $\overrightarrow{\phi}$ can be written as
a $2N\left(  2l_{\max}\mathtt{+}1\right)  \mathtt{\times}1$ vector which
imposes informations of incident waves (see details in Appendix B).

At this stage, we should point out that the above equations for multiple
magnetic scattering of massless Dirac quasiparticles are similar to those for
multiple nonmagnetic scattering of massive Dirac quasiparticles, but totally
different from those for multiple nonmagnetic scattering of massless Dirac
quasiparticles, since the expressions therein can be simplified as a more
compact form due to $s_{-\left(  l+1\right)  }^{\prime}=s_{l}^{\prime}$.

The above theory enables to solve multiple magnetic scattering problems in
gapless TI surfaces with higher partial waves, which could be important as
distances between scatterers decrease or the scattering potential is
strengthened. One simple application is to calculate the magnetic scattering
CSs. To calculate the CSs, we have to take the approximations $H_{l}%
^{(1)}\left(  z\right)  \rightarrow\sqrt{\frac{2}{\pi z}}e^{i\left(
z-\frac{l\pi}{2}-\frac{\pi}{4}\right)  }$, and $e^{il\theta_{n}}=\left[
\frac{\boldsymbol{\rho}_{n}\cdot\left(  \hat{x}+i\hat{y}\right)  }{\rho_{n}%
}\right]  ^{l}\approx\left[  \frac{\boldsymbol{r}\cdot\left(  \hat{x}+i\hat
{y}\right)  }{r}\right]  ^{l}=e^{il\varphi}$ for large distance. As a result,
\begin{equation}
\Psi_{\text{sc}}\left(  \boldsymbol{r}\right)  \rightarrow f\left(
\boldsymbol{k},\varphi\right)  \frac{e^{ikr}}{\sqrt{2r}}\left(
\begin{array}
[c]{c}%
1\\
\mp ie^{i\varphi}%
\end{array}
\right)  ,
\end{equation}
where $f\left(  \boldsymbol{k},\varphi\right)  $ is the scattering amplitude,
from which we have the differential and total CSs as follows:
\begin{align}
\frac{d\Lambda}{d\varphi}  &  =\left\vert f\left(  \boldsymbol{k}%
,\varphi\right)  \right\vert ^{2},\\
\Lambda_{tot}  &  =\int_{0}^{2\pi}d\varphi\left\vert f\left(  \boldsymbol{k}%
,\varphi\right)  \right\vert ^{2}=\sqrt{\frac{8\pi}{k}}\operatorname{Im}%
\left[  e^{-i\pi/4}f\left(  \boldsymbol{k},\varphi=0\right)  \right]  .
\label{ot}%
\end{align}
Here, we have used the two-dimensional optical theorem.

Besides, we could also obtain the transverse component of resistivity (or say
the analog of Hall component in the case with magnetic field)%
\begin{equation}
\Omega=\int_{0}^{2\pi}d\varphi\left\vert f\left(  \boldsymbol{k}%
,\varphi\right)  \right\vert ^{2}\sin\varphi, \label{Hall}%
\end{equation}
and the inverse electron momentum relaxation time (the quantity proportional
to the dissipative component of resistivity)%
\begin{equation}
\Gamma_{M}=\int_{0}^{2\pi}d\varphi\left\vert f\left(  \boldsymbol{k}%
,\varphi\right)  \right\vert ^{2}\left(  1-\cos\varphi\right)  .
\end{equation}

\section{Results and discussions}

In the following calculations, without losing the general properties, we shall
just consider the incident wave $\Phi_{+}^{in}\mathtt{=}\frac{e^{ikx}}%
{\sqrt{2}}(%
\begin{array}
[c]{cc}%
1\mathtt{,} & \mathtt{-}i
\end{array}
)^{\text{T}}$ propagating along the positive $\hat{\boldsymbol{x}}$ direction.
In particular, for a single magnetic impurity scattering, we can obtain the
scattering amplitude including all of the partial waves, which is written as
\begin{equation}
f\left(  \boldsymbol{k},\varphi\right)  =\left\{  f_{0}\left(  \varphi\right)
+\sum_{l=1}^{\infty}\left[  f_{l}\left(  \varphi\right)  +f_{-l}\left(
\varphi\right)  \right]  \right\}  e^{i\left(  \boldsymbol{k}-k\hat{r}\right)
\cdot\boldsymbol{r}^{\prime}}\label{amp}%
\end{equation}
with $f_{0}\left(  \varphi\right)  =\sqrt{\frac{2}{i\pi k}}s_{0}$ and $f_{\pm
l}\left(  \varphi\right)  =\sqrt{\frac{2}{i\pi k}}s_{\pm l}e^{\pm il\varphi}$.
The differential and total CSs are given by%
\begin{align}
\frac{d\Lambda}{d\varphi} &  =\frac{2}{\pi k}\left\vert \left\{  s_{0}%
+\sum_{l=1}^{\infty}\left[  s_{l}e^{il\varphi}+s_{-l}e^{-il\varphi}\right]
\right\}  \right\vert ^{2},\label{dcs}\\
\Lambda_{tot} &  =-\frac{4}{k}\left[  \operatorname{Re}\left(  s_{0}\right)
+\sum_{l=1}^{\infty}\operatorname{Re}\left(  s_{l}+s_{-l}\right)  \right]
.\label{tcs}%
\end{align}
This total CS equation is obtained from the optical theorem. It is clear that
the $s-$wave is independent on the direction of scattered wave, therefore, we
have to introduce higher partial waves, such as $p\mathtt{-}$ and
$d\mathtt{-}$waves, and so on. If one just considers the $s-$wave in
calculations, differential CS may lead to an unreasonable result of $\left.
d\Lambda/d\varphi\right\vert _{\varphi=\pi}=0$ (backscattering is forbidden)
for some particular effective magnetic moment $M$. This is different from the
nonmagnetic impurity scattering on TI surface \cite{ZGFu2011} as well as on
conventional 2DEG with weak Rashba SOC \cite{Walls2006}, where the
$s\mathtt{-}$wave approximation should be a reasonable
choice.\begin{figure}[ptb]
\begin{center}
\includegraphics[width=.7\linewidth]{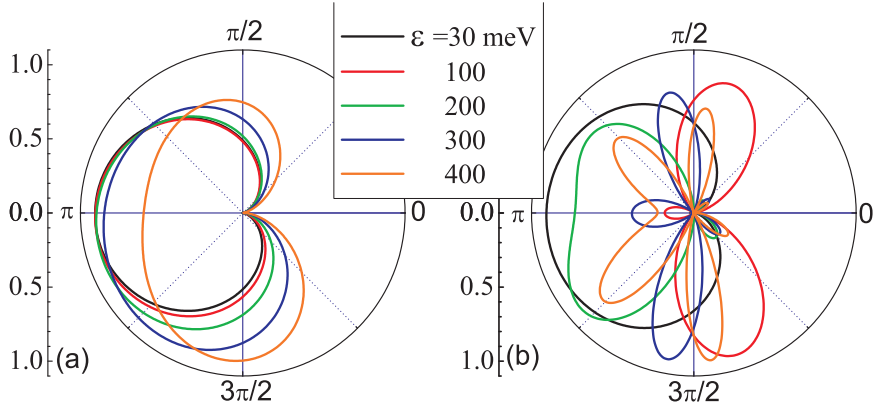}
\end{center}
\caption{ (Color online) The normalized differential CSs $d\Lambda/d\phi$ for
(a) a single and (b) two magnetic scatterers located at $\boldsymbol{r}%
_{1,2}=(\pm3,0)$ on TI surface with effective magnetic moment $M=60$ meV. The
radius of scatterer $a\mathtt{=}1$ nm, and $l_{\max}=2$ are chosen.}%
\label{fig1}%
\end{figure}

The results of normalized differential CS as a function of energy $\epsilon$
for the massless Dirac electron scattered by a single magnetic impurity
absorbed on TI surface with effective magnetic moment $M\mathtt{=}60$ meV are
shown in Fig. \ref{fig1}(a). In the calculations we take $l_{\max}=2$, which
works out convergent results. Different from the nonmagnetic impurity
scattering case, the backscattering is obvious in the differential CS, i.e.,
$\left.  d\Lambda/d\varphi\right\vert _{\varphi=\pi}\neq0$, in present case
since the time-reversal symmetry is broken by magnetic impurity scattering.
Furthermore, we find that for the weak effective magnetic moment $M$ (such as
the values chosen in this work $M\leq100$ meV), the backscattering is greater
than forward scattering. However, if the effective magnetic moment is large
enough (for example, when $M\sim500$ meV and $\epsilon=450$ meV) we find the
backscattering is weaker than the forward scattering (not shown here).

Two-impurity scattering provides a good test-bed to highlight the coexistence
of various scattering phenomena, including transmission, reflection,
interference, and resonance. The corresponding CS offers a measure of
interaction events between the two impurity centers, and interference effects
are useful in revealing actual electron density currents on TI surfaces. For
instance, if two impurities are close to each other, the electronic
wavefunctions will be scattered from both impurities, resulting in quantum
interference. From the above theory, we can obtain the simple expression of
scattering amplitude just containing the $s\mathtt{-}$wave ($l_{\max}=0$),
which is given by%
\begin{align}
f_{0}^{N}\left(  \boldsymbol{k},\varphi\right)   &  =\sum_{n,m=1}^{N}%
\frac{s_{0}e^{i\left(  \boldsymbol{k}\cdot\boldsymbol{r}_{m}-k\hat{r}%
\cdot\boldsymbol{r}_{n}\right)  }}{\sqrt{2i\pi k}}\\
&  \times\left\{  \left[  D^{-1}\right]  _{\left(  2n-1\right)  ,\left(
2m-1\right)  }+\left[  D^{-1}\right]  _{2n,2m}\right. \nonumber\\
&  \left.  \pm\left[  \left[  D^{-1}\right]  _{2n,\left(  2m-1\right)
}+\left[  D^{-1}\right]  _{\left(  2n-1\right)  ,2m}\right]  \right\}
.\nonumber
\end{align}
\begin{figure}[ptb]
\begin{center}
\includegraphics[width=.7\linewidth]{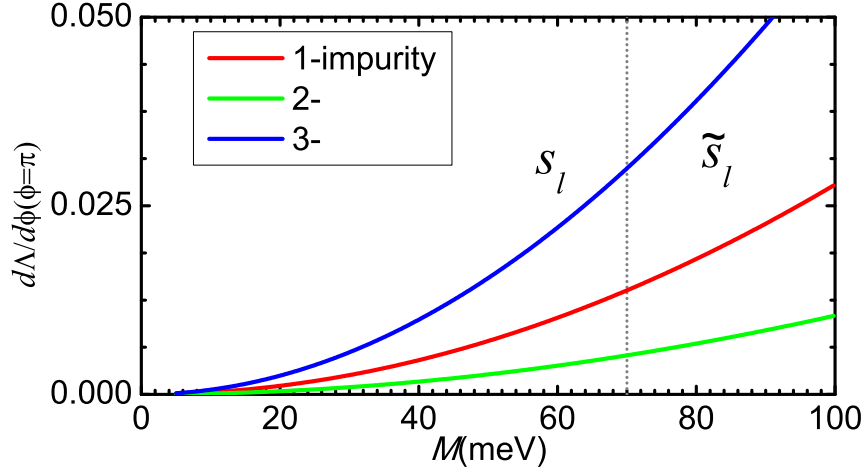}
\end{center}
\caption{ (Color online) The magnetic moment $M$ dependence of differential CS
$d\Lambda/d\phi$ along the negative direction of $x$ axis $\phi=\pi$ for one
magnetic scatterer (red line), two scatterers (green line) located at
$\boldsymbol{r}_{1,2}=(\pm3,0)$, and three scatterers (blue) located at
$\boldsymbol{r}_{1}=(-3,0)$, $\boldsymbol{r}_{2}=(3,-1)$, $\boldsymbol{r}%
_{3}=(2,4)$. The Fermi energy is chosen as $\epsilon=70$ meV.}%
\label{fig2}%
\end{figure}However, if higher partial waves ($l_{\max}\geq1$) are taken into
account, the scattering amplitude expression for $N$ magnetic impurities
becomes tedious and complex since $D-$matix becomes a lager one.

Typical numerical results of differential CSs for two magnetic impurities
located at $\boldsymbol{r}_{1,2}=\left(  \pm3,0\right)  $ are presented in
Fig. \ref{fig1}(b). We also note that the interference effect is related not
only to the effective magnetic moment $M$ but also to the configuration of
impurities. For the present considered impurity locations $\boldsymbol{r}%
_{1,2}$, on one hand, we find from Fig. \ref{fig1}(b) that the backscattering
is more prominent than the forward scattering. On the other hand, comparing
with the nonmagnetic double-impurity scattering on gapless TI surface, the
symmetry of differential CSs for two identical magnetic scatterers is reduced.

On one hand, independent on the impurity locations, with increasing the
effective magnetic moment $M$, we find that the relative strength of
backscattering becomes more and more remarkable since the differential CS
along the negative $\hat{\boldsymbol{x}}$ direction $\left.  d\Lambda
/d\varphi\right\vert _{\varphi=\pi}$ increases with $M$, see Fig. \ref{fig2}
with the Fermi energy $\epsilon=70$ meV. On the other hand, comparing with the
scattering from a single (red curve) magnetic impurity, two (green curve) or
three (blue curve) impurities will weaken or strengthen the backscattering due
to interference effect, which is dependent on the impurity configurations
relative to the direction of incident wave. We must point out that in the
calculations, we should use $\tilde{s}_{l}$ for the energy region of
$\left\vert \epsilon\right\vert <M$, while $s_{l}$ for the energy region of
$\left\vert \epsilon\right\vert >M$, which are denoted in Fig. \ref{fig2}.
\begin{figure}[ptb]
\begin{center}
\includegraphics[width=.7\linewidth]{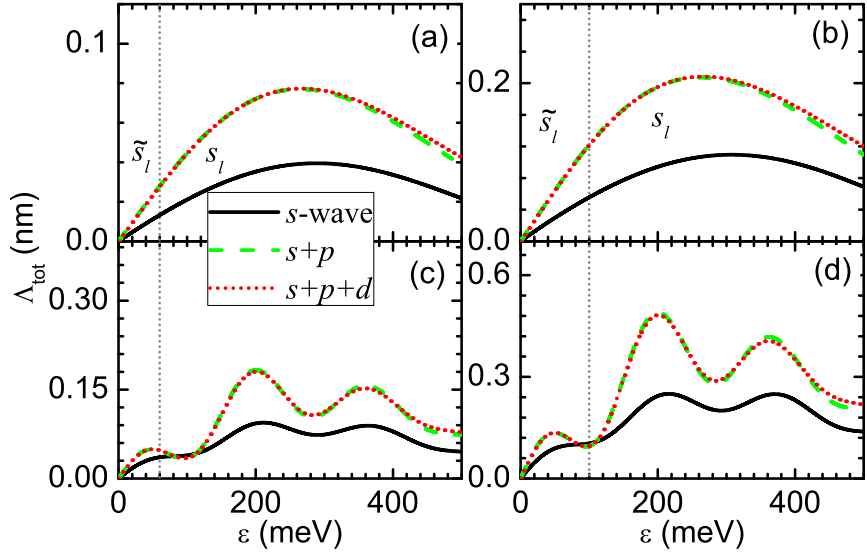}
\end{center}
\caption{ (Color online) The total CS $\Lambda_{tot}$ for a single (a-b) and
two (c-d) magnetic impurities located at $\boldsymbol{r}_{1,2}=(\pm3,0)$. The
effective magnetic moment is chosen as $M=60$ meV in (a)and (c), while $M=100$
meV in (b) and (d), respectively.}%
\label{fig3}%
\end{figure}

Now let us turn to discuss the total CSs, which are exhibited in Fig.
\ref{fig3}. As mentioned above, the $s\mathtt{-}$wave approximation cannot
give out convergent result, see the black curves in Fig. \ref{fig3}, whereas,
when we introduce higher partial waves (such as $l_{\max}=2$ chosen in our
calculations), the total CSs becomes convergent ultimately. Differing from the
nonmagnetic impurity scattering, although the higher partial waves can induce
remarkable corrections, we have not found additional resonant peaks in total
CSs due to higher partial waves. Moreover, it is obvious that interference
between double impurities brings about oscillations in total CSs (see Figs.
\ref{fig3}(c) and \ref{fig3}(d)), which cannot be observed in the case of
nonmagnetic impurity scattering \cite{ZGFu2011}. From numerical calculations,
we find on one hand that, the optical theorem is correct and should
characterize the general multiple-scattering processes, since the results
obtained by the numerical integration of the first equality in Eq. (\ref{ot})
are in good agreement with that obtained from the second equality; On the
other hand, the curves of total CSs are smooth at the energy of $\epsilon=M$,
which indicates that the limiting function of $s_{l}$ and $\tilde{s}_{l}$,
i.e., Eq. (\ref{limit}) is reasonable. Besides, for the much strong effective
magnetic impurity scattering, with increasing the Fermi energy we find that
the total CS for single- (double-) impurity is convergent to 4 nm (8 nm). This
is also different from the nonmagnetic impurity scattering, where the total
CSs converge to zero with increasing the energy of Dirac electrons
\cite{ZGFu2011}. \begin{figure}[ptb]
\begin{center}
\includegraphics[width=.7\linewidth]{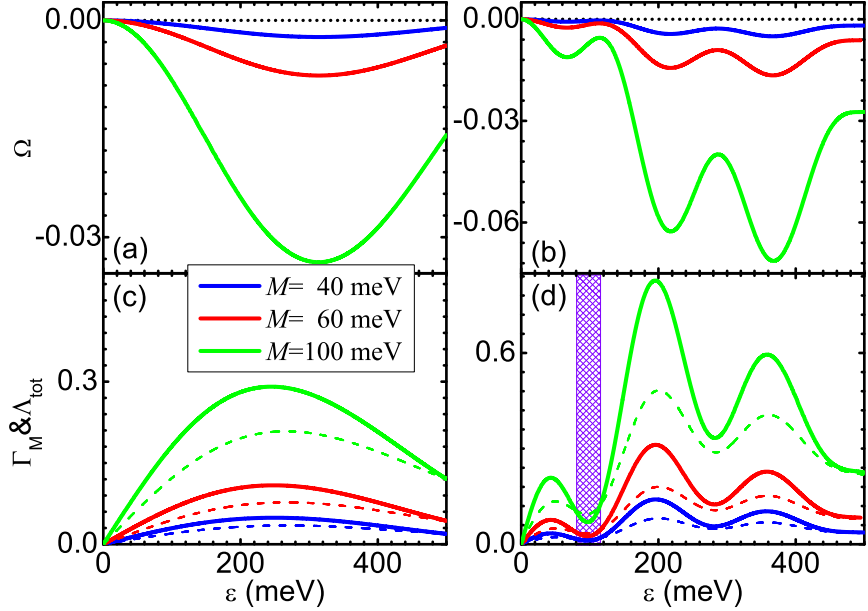}
\end{center}
\caption{ (Color online) Hall component of resistivity $\Omega$ for a single
(a) and two (b) magnetic impurities on TI surface; Inverse momentum relation
time $\Gamma_{M}$ (thick solid curves) and total CS $\Lambda_{tot}$ (thin
dashed curved) for a single (c) and two (d) magnetic impurities as functions
of energy $\epsilon$.}%
\label{fig4}%
\end{figure}

In spite of the differential and total CSs, we also calculated the transverse
component of resistivity $\Omega$, which is analogous to Hall component in the
case with external magnetic field. The typical results of $\Omega$ as a
function of $\epsilon$ for single- and double-impurity with different $M$ are
listed in Figs. \ref{fig4}(a) and \ref{fig4}(b), respectively. We find that
the Hall component of the resistivity $\Omega$ always keeps its sign as
negative (i.e., $\Omega<0$), which is independent on the impurity locations.
In the numerical calculations, we take $l_{\max}\geq2$ which results in a
convergent result, however, taking into account higher partial waves, it is
difficult to be obtained analytically from Eqs. (\ref{Hall}) and (\ref{amp})
since the expression for integral result is tedious and complex. Therefore,
the low-energy Dirac electrons are deflected to one side of TI sample due to
magnetic impurity scattering. This fact may be helpful for tuning the Hall
voltage of sample. Interestingly, we note that this type of Hall component
also occurs in the nonmagnetic impurity scattering on gapless TI surface. We
now switch gears and consider the case of double magnetic impurities on TI
surface again. Different from the single case, $\Omega$ exhibits oscillating
behavior due to the interference during the multiple impurities scattering
processes, as shown in Fig. \ref{fig4}(b).

Before ending this paper, we would like to discuss another important quantity,
the inverse electron momentum relaxation time $\Gamma_{M}$, which is
proportional to the dissipative component of resistivity. The numerical
results are plotted in Figs. \ref{fig4}(c) and \ref{fig4}(d) for single- and
double-impurity cases, respectively. The behavior of $\Gamma_{M}$ (thick solid
curves) is qualitatively similar to the one for the total CSs $\Lambda_{tot}$
(thin dashed curves), and the interference effect is also clear in the
$\Gamma_{M}$ induced by double-impurity, see Fig. \ref{fig4}(d). The inverse
electron momentum relaxation time $\Gamma_{M}$ is a fairly sensitive quantity
that determines whether the charge carrier is attracted to an impurity or is
repelled from it. It is also useful for determining whether the backscattering
is greater than the forward scattering by comparing it with the total CSs
(note that both $\Lambda_{tot}$ and $\Gamma_{M}$ have the dimension of an
length in two dimensional scattering). If $\Lambda_{tot}\mathtt{<}\Gamma_{M}$
($\Lambda_{tot}\mathtt{>}\Gamma_{M}$), there is more (less) backscattering
than forward scattering. Taking $M=60$ meV as an example, we find
$\Lambda_{tot}\mathtt{<}$ $\Gamma_{M}$, as revealed by the red curves in Fig.
\ref{fig4}(c), which indicates that the backscattering is greater than the
forward scattering in the low-energy region. This fact is consistent with the
behavior of differential CSs shown in Fig. \ref{fig1}(a). Particularly, by
observing the green curves in Fig. \ref{fig4}(d) of energy region of
$80\mathtt{\sim}120$ meV (the shadow region), one can find $\Lambda
_{tot}\mathtt{>}\Gamma_{M}$. This suggests that due to the interference the
backscattering is weaker than forward scattering in this energy region for the
present double magnetic impurity locations, which is also consistent with the
analysis of differential CSs. Consequently, we believe that our results shown
here are reasonable, and we hope our findings could be detected in the future experiments.

\section{Conclusions}

In summary, we have proposed a general low-energy multiple-scattering
partial-wave theory for quasiparticles on the gapless topological insulator
(TI) surfaces in the presence of magnetic impurities. Based on this theory,
one can solve the scattering problems of $N$ magnetic impurities. As an
application, we have calculated the CSs, the inverse momentum relaxation time,
and the transverse resistivity component for a single and two circular
magnetic scattering. We have found that the usual $s\mathtt{-}$wave
approximation is not sufficient, while higher partial waves must be introduced
to obtain convergent results. On the gapless TI surfaces, differing from the
single nonmagnetic impurity case, the backscattering occurs and becomes much
stronger with increasing the effective magnetic moment $M$. Interference
effects are obvious in CSs from quasiparticle scattering off two magnetic
scattering centers, and oscillating behaviors are introduced in $\Lambda
_{tot}$ associated with higher-order partial-waves. A non-zero perpendicular
resistivity component has also been shown. Similar to the total CS, the
inverse momentum relaxation time and the transverse resistivity component
exihibit\ oscillations for multiple magnetic scattering centers due to
interference. Furthermore, our theory could be extended to spin-polarized
case. It could also be applied to simulate the electron flow (charge current
and spin current) through a quantum point contact on a TI surface by
monitoring the changes in conductance through the quantum point contact as a
moveable STM tip is scanned above the surface of TI.

\begin{acknowledgments}
This work was supported by NSFC under Grants No. 90921003, No. 60776063, and
No. 60821061, and by the National Basic Research Program of China (973
Program) under Grants No. 2009CB929103 and No. G2009CB929300.
\end{acknowledgments}


\section*{APPENDIX A: DERIVATION DETAILS OF SCATTERED WAVEFUNCTION}

Starting from the model considered in the main text, one can find the spinor
spherical wavefunctions inside the magnetic scattering potential ($r<a$) as
follows:
\begin{align}
\zeta_{l}^{(1,2)}\left(  \boldsymbol{r},k^{\prime},\pm\right)   &
=\frac{we^{il\theta}}{\sqrt{2\left\vert \epsilon\right\vert k^{\prime}}%
}\left(
\begin{array}
[c]{c}%
\sqrt{\left\vert \epsilon+M\right\vert }H_{l}^{(1,2)}\left(  k^{\prime
}r\right) \\
\pm\sqrt{\left\vert \epsilon-M\right\vert }H_{l+1}^{(1,2)}\left(  k^{\prime
}r\right)  e^{i\theta}%
\end{array}
\right)  ,\text{ \ \ }(\left\vert \epsilon\right\vert >M),\tag{A1}\\
\widetilde{\zeta}_{l}\left(  \boldsymbol{r},k^{\prime},\pm\right)   &
=\frac{we^{il\theta}}{\sqrt{2\left\vert \epsilon\right\vert k^{\prime}}%
}\left(
\begin{array}
[c]{c}%
\sqrt{\left\vert \epsilon+M\right\vert }I_{l}\left(  k^{\prime}r\right) \\
\mp\sqrt{\left\vert \epsilon-M\right\vert }I_{l+1}\left(  k^{\prime}r\right)
e^{i\theta}%
\end{array}
\right)  ,\text{ \ \ }(\left\vert \epsilon\right\vert <M), \tag{A2}%
\end{align}
where $k^{\prime}=\frac{\sqrt{\left\vert \epsilon^{2}-M^{2}\right\vert }%
}{\hbar v_{f}}$ and $w=e^{i\phi}$ is an overall phase meaning of the
nonrelativistic wavefunction in the rest frame of $\epsilon=M$. Here,
$H_{l}^{(1,2)}\left(  z\right)  $ and $I_{l}\left(  z\right)  $ are $l^{th}%
$-order Hankel functions and modified Bessel functions of first kind,
respectively. We would like to point out that since the wavefunctions should
be zero at $r=0$, we have neglected the modified Bessel functions of second
kind $K_{l}\left(  z\right)  $ for $\left\vert \epsilon\right\vert <M$, which
are emanative in the limit of $z\rightarrow0$. Whereas, the wavefunctions
outside the magnetic potential ($r>a$) can be expressed as
\begin{equation}
\chi_{l,\pm}^{(1,2)}\left(  \boldsymbol{r}\right)  =\frac{1}{\sqrt{2k}}\left(
\begin{array}
[c]{c}%
H_{l}^{(1,2)}\left(  kr\right)  e^{il\theta}\\
\pm H_{l+1}^{(1,2)}\left(  kr\right)  e^{i\left(  l+1\right)  \theta}%
\end{array}
\right)  , \tag{A3}%
\end{equation}
with $k=\frac{\epsilon}{\hbar v_{f}}$. $\chi^{(1)}$ ($\chi^{(2)}$) denotes the
outgoing (incoming) cylindrical wave about $\boldsymbol{r}$=$0$. The incident
plane-wave centered about a single scatterer located at $\boldsymbol{r}_{n}$
is given by
\begin{equation}
\Phi_{\pm}^{in}\left(  \boldsymbol{r}\right)  \mathtt{=}\sum_{l=-\infty
}^{\infty}\frac{\sqrt{k}e^{i\boldsymbol{k}\cdot\boldsymbol{r}_{n}}e^{il\left(
\theta_{n}-\theta_{\boldsymbol{k}}\right)  }}{2}i^{l}\left(  \chi_{l,\pm
}^{(1)}\left(  \boldsymbol{\rho}_{n}\right)  \mathtt{+}\chi_{l,\pm}%
^{(2)}\left(  \boldsymbol{\rho}_{n}\right)  \right)  , \tag{A4}%
\end{equation}
where $\boldsymbol{\rho}_{n}\mathtt{=}\boldsymbol{r}\mathtt{-}\boldsymbol{r}%
_{n}$ and $e^{i\theta_{n}}\mathtt{=}\frac{\boldsymbol{\rho}_{n}\cdot\left(
\hat{x}+i\hat{y}\right)  }{\rho_{n}}$. Remember that $\theta_{\boldsymbol{k}}$
is the angle defining the direction of the wave vector, and $\theta_{n}$ is
related to the direction of $\boldsymbol{r}\mathtt{-}\boldsymbol{r}_{n}$ in
this equation. Then, the fully scattered wave function in the region $\rho
_{n}>a$ can be written as explicitly
\begin{align}
\Psi_{\text{I}}\left(  \boldsymbol{r},\boldsymbol{k},\pm\right)   &
=\Phi_{\pm}^{in}\left(  \boldsymbol{r}\right)  +\Psi_{\text{sc}}\left(
\boldsymbol{r},\boldsymbol{k},\pm\right) \nonumber\\
&  =\sqrt{k}e^{i\boldsymbol{k}\cdot\boldsymbol{r}_{n}}\sum_{l=-\infty}%
^{\infty}i^{l}\left[  \frac{1}{2}e^{2i\delta_{l}}\chi_{l}^{\left(  1\right)
}\left(  \boldsymbol{\rho}_{n},\boldsymbol{k},\pm\right)  +\frac{1}{2}\chi
_{l}^{\left(  2\right)  }\left(  \boldsymbol{\rho}_{n},\boldsymbol{k}%
,\pm\right)  \right]  e^{il\left(  \theta_{n}-\theta_{\boldsymbol{k}}\right)
}, \tag{A5}%
\end{align}
where $\delta_{l}$ are phase shifts of the outgoing cylindrical partial waves,
$\chi_{l}^{\left(  1\right)  }\left(  \boldsymbol{\rho}_{n},\boldsymbol{k}%
,\pm\right)  $. In the region of $\rho_{n}\leq a$, the fully scattered wave
function is given by
\begin{equation}
\Psi_{\text{II}}\left(  \boldsymbol{r},\boldsymbol{k^{\prime}},\pm\right)
=\left\{
\begin{array}
[c]{c}%
\sqrt{k^{\prime}}e^{i\boldsymbol{k}\cdot\boldsymbol{r}_{n}}\sum_{l=-\infty
}^{\infty}i^{l}d_{l}\left[  \frac{1}{2}\zeta_{l}^{\left(  1\right)  }\left(
\boldsymbol{\rho}_{n},\boldsymbol{k^{\prime}},\pm\right)  -\frac{1}{2}%
\zeta_{l}^{\left(  2\right)  }\left(  \boldsymbol{\rho}_{n}%
,\boldsymbol{k^{\prime}},\pm\right)  \right]  e^{il\left(  \theta_{n}%
-\theta_{\boldsymbol{k}}\right)  },\text{ \ \ }\left(  \left\vert
\epsilon\right\vert >M\right) \\
\sqrt{k^{\prime}}e^{i\boldsymbol{k}\cdot\boldsymbol{r}_{n}}\sum_{l=-\infty
}^{\infty}i^{l}\widetilde{d}_{l}\widetilde{\zeta}_{l}\left(  \boldsymbol{r}%
,k^{\prime},\pm\right)  e^{il\left(  \theta_{n}-\theta_{\boldsymbol{k}%
}\right)  },\text{ \ \ }\left(  \left\vert \epsilon\right\vert <M\right)
\end{array}
\right.  . \tag{A6}%
\end{equation}
By the continuity of the wavefunction at $\rho_{n}=a$, $\Psi_{\text{I}}\left(
\boldsymbol{a},\boldsymbol{k},\pm\right)  =\Psi_{\text{II}}\left(
\boldsymbol{a},\boldsymbol{k}^{\prime},\pm\right)  $, we can obtain the
scattered wavefunction Eq. (\ref{scw}), and the scattering amplitude $s_{l}$
and $\widetilde{s}_{l}$ shown in Eqs. (\ref{ss}) and (\ref{ss2}) in main text.
It is clear that the scattered wavefunctions shown in Eq. (\ref{scw}) are
different from those for nonmagnetic impurity scattering on gapless TI
surface, where
\begin{equation}
\Psi_{\text{sc}}\left(  \boldsymbol{r}\right)  \mathtt{=}\sum_{l=0}^{\infty
}\frac{4i\hbar v_{f}s_{l}^{\prime}}{k}G_{l}\left(  \boldsymbol{r}%
,\boldsymbol{r}_{n},\epsilon\right)  T_{l}^{^{\prime}}\left[  \Phi_{\pm}%
^{in}\right]  \tag{A7}%
\end{equation}
with
\begin{equation}
G_{l}\mathtt{\propto}\left(
\begin{array}
[c]{cc}%
H_{l}^{(1)}e^{il\theta_{n}} & \mp H_{l+1}^{(1)}e^{-i\left(  l+1\right)
\theta_{n}}\\
\pm H_{l+1}^{(1)}e^{i\left(  l+1\right)  \theta_{n}} & H_{l}^{(1)}%
e^{-il\theta_{n}}%
\end{array}
\right)  , \tag{A8}%
\end{equation}
$T_{l}^{^{\prime}}\mathtt{=}\mathtt{diag}(%
\begin{array}
[c]{cc}%
\hat{P}_{l}^{-}, & \hat{P}_{l}^{+}%
\end{array}
)$, and
\begin{equation}
s_{l}^{\prime}=\frac{J_{l}\left(  \kappa^{\prime}a\right)  J_{l+1}\left(
ka\right)  -J_{l}\left(  ka\right)  J_{l+1}\left(  \kappa^{\prime}a\right)
}{H_{l}^{(1)}\left(  ka\right)  J_{l+1}\left(  \kappa^{\prime}a\right)
-H_{l+1}^{(1)}\left(  ka\right)  J_{l}\left(  \kappa^{\prime}a\right)  }.
\tag{A9}%
\end{equation}
Here $\kappa^{\prime}\mathtt{=}\frac{\epsilon-V_{0}}{\hbar v_{f}}$, and
$V_{0}$ is the scalar potential.

\subsection*{APPENDIX B: DETAILS OF THE EXPEDITION FOR MULTIPLE SCATTERED
WAVE}

In order to understand the extending operation, we would like to start from
the $s\mathtt{-}$wave scattering for two identical magnetic impurities. The
total wavefunction can be written as
\begin{equation}
\Psi\left(  \boldsymbol{r}\right)  =\Phi\left(  \boldsymbol{r}\right)
+\sum_{n=1}^{2}s_{0}^{\left(  n\right)  }G_{+0}\left(  \boldsymbol{r}%
,\boldsymbol{r}_{n},\epsilon\right)  T_{0}^{-}\left[  \Psi_{n}\left(
\boldsymbol{r}_{n},\boldsymbol{k},\pm\right)  \right]  , \tag{B1}%
\end{equation}
where%
\begin{equation}
\Psi_{n}\left(  \boldsymbol{r}\right)  =\Phi\left(  \boldsymbol{r}\right)
+\sum_{m\neq n}^{2}s_{0}^{\left(  m\right)  }G_{+0}\left(  \boldsymbol{r}%
,\boldsymbol{r}_{m},\epsilon\right)  T_{0}^{-}\left[  \Psi_{m}\left(
\boldsymbol{r}_{m},\boldsymbol{k},\pm\right)  \right]  . \tag{B2}%
\end{equation}
Equation (B1) indicates that if the value of $\Psi\left(  \boldsymbol{r}%
\right)  $ and its derivatives due to the $\hat{P}_{0,1}^{\pm}$ dependence of
$T_{0}^{\pm}$ at each scatterer is known, the entire wavefunction $\Psi\left(
\boldsymbol{r}\right)  $ is completely determined. We can calculate the
derivatives of $\Psi_{1}$ at $\boldsymbol{r}_{1}$ and $\Psi_{2}$ at
$\boldsymbol{r}_{2}$ and combine the result into a matrix equation, which are
given by
\begin{equation}
\left(
\begin{array}
[c]{c}%
T_{0}^{-}\left[  \Psi_{1}\left(  \boldsymbol{r}_{1}\right)  \right] \\
T_{0}^{-}\left[  \Psi_{2}\left(  \boldsymbol{r}_{2}\right)  \right]
\end{array}
\right)  =D^{-1}\left(
\begin{array}
[c]{c}%
T_{0}^{-}\left[  \Phi\left(  \boldsymbol{r}_{1}\right)  \right] \\
T_{0}^{-}\left[  \Phi\left(  \boldsymbol{r}_{2}\right)  \right]
\end{array}
\right)  , \tag{B3}%
\end{equation}
where
\begin{equation}
D=\mathbf{1}_{4\times4}-\mathbf{G}_{4\times4}\mathbf{S}_{4\times4}, \tag{B4}%
\end{equation}
with%
\begin{align}
\mathbf{G}  &  \mathbf{=}\left(
\begin{array}
[c]{cc}%
\mathbf{0} & T_{0}^{-}\left[  G_{+0}\left(  \boldsymbol{r}_{1},\boldsymbol{r}%
_{2},\epsilon\right)  \right] \\
T_{0}^{-}\left[  G_{+0}\left(  \boldsymbol{r}_{2},\boldsymbol{r}_{1}%
,\epsilon\right)  \right]  & \mathbf{0}%
\end{array}
\right)  ,\tag{B5}\\
\mathbf{S}  &  \mathbf{=}\left(
\begin{array}
[c]{cc}%
s_{0}^{\left(  1\right)  } & 0\\
0 & s_{0}^{\left(  2\right)  }%
\end{array}
\right)  \otimes\mathbf{1}_{2\times2}. \tag{B6}%
\end{align}
Finally, the total wavefunction is written as
\begin{equation}
\Psi\left(  \boldsymbol{r}\right)  \mathtt{=}\Phi_{\pm}^{in}\mathtt{+}%
\mathbb{G}\left(  \boldsymbol{r}\right)  \mathbf{S}D^{-1}\left(
\begin{array}
[c]{c}%
T_{0}^{-}\left[  \Phi\left(  \boldsymbol{r}_{1}\right)  \right] \\
T_{0}^{-}\left[  \Phi\left(  \boldsymbol{r}_{2}\right)  \right]
\end{array}
\right)  \tag{B7}%
\end{equation}
with
\begin{equation}
\mathbb{G}\left(  \boldsymbol{r}\right)  =\left(
\begin{array}
[c]{cc}%
G_{+0}\left(  \boldsymbol{r},\boldsymbol{r}_{1}\right)  , & G_{+0}\left(
\boldsymbol{r},\boldsymbol{r}_{2}\right)
\end{array}
\right)  . \tag{B8}%
\end{equation}
Taking into account $l_{\max}\geq1$ partial waves, for $N$ magnetic scatterers
located at positions $\boldsymbol{r}_{1},\boldsymbol{r}_{2},\cdots
\boldsymbol{r}_{N}$, the scattered wavefunction is given by Eq. (\ref{10}) in
main text. For numerical calculations, we have to align the matrix elements
reasonably, thereby, we define symbols $\mu_{0}=2\left(  n-1\right)  \left(
2l_{\max}+1\right)  +1$, $\lambda_{0}=\mu_{0}+1$, $\alpha_{0}=2\left(
n-1\right)  \left(  2l_{\max}+1\right)  +4l-1$, $\gamma_{0}=\alpha_{0}+1$,
$\tau_{0}=2\left(  n-1\right)  \left(  2l_{\max}+1\right)  +4l+1$, $\eta
_{0}=\tau_{0}+1$, $\nu_{0}=2\left(  m-1\right)  \left(  2l_{\max}+1\right)
+1$, $\beta_{0}=\nu_{0}+1$, $\alpha=2\left(  m-1\right)  \left(  2l_{\max
}+1\right)  +4l^{\prime}-1$, $\gamma=\alpha+1$, $\nu=2\left(  m-1\right)
\left(  2l_{\max}+1\right)  +4l^{\prime}+1$, and $\beta=\nu+1$. Following this
way $\mathbb{G}\left(  \boldsymbol{r}\right)  $ (a $2\mathtt{\times}2N\left(
2l_{\max}\mathtt{+}1\right)  $ matrix) is aligned as
\begin{equation}
\mathbb{G}\left(  \boldsymbol{r}\right)  =\left(
\begin{array}
[c]{ccc}%
\widetilde{G}\left(  \boldsymbol{r},\boldsymbol{r}_{1}\right)  , &
\widetilde{G}\left(  \boldsymbol{r},\boldsymbol{r}_{2}\right)  , & \cdots,
\end{array}%
\begin{array}
[c]{c}%
\widetilde{G}\left(  \boldsymbol{r},\boldsymbol{r}_{N}\right)
\end{array}
\right)  \tag{B9}%
\end{equation}
with $\widetilde{G}\left(  \boldsymbol{r},\boldsymbol{r}_{i}\right)
\mathtt{=}[%
\begin{array}
[c]{ccc}%
G_{+0}, & G_{+1}, & G_{-1},
\end{array}
\cdots,%
\begin{array}
[c]{cc}%
G_{+l_{\max}}, & G_{-l_{\max}}%
\end{array}
]$. Explicitly, for $l=0$, we align%
\begin{equation}
\left(
\begin{array}
[c]{cc}%
\mathbb{G}\left(  \boldsymbol{r}\right)  _{1,\mu_{0}} & \mathbb{G}\left(
\boldsymbol{r}\right)  _{1,\lambda_{0}}\\
\mathbb{G}\left(  \boldsymbol{r}\right)  _{2,\mu_{0}} & \mathbb{G}\left(
\boldsymbol{r}\right)  _{2,\lambda_{0}}%
\end{array}
\right)  =G_{+0}\left(  \boldsymbol{r},\boldsymbol{r}_{n}\right)  , \tag{B10}%
\end{equation}
and for $l\geq1$,%
\begin{align}
\left(
\begin{array}
[c]{cc}%
\mathbb{G}\left(  \boldsymbol{r}\right)  _{1,\alpha_{0}} & \mathbb{G}\left(
\boldsymbol{r}\right)  _{1,\gamma_{0}}\\
\mathbb{G}\left(  \boldsymbol{r}\right)  _{2,\alpha_{0}} & \mathbb{G}\left(
\boldsymbol{r}\right)  _{2,\gamma_{0}}%
\end{array}
\right)   &  =G_{+l}\left(  \boldsymbol{r},\boldsymbol{r}_{n}\right)
,\tag{B11}\\
\left(
\begin{array}
[c]{cc}%
\mathbb{G}\left(  \boldsymbol{r}\right)  _{1,\tau_{0}} & \mathbb{G}\left(
\boldsymbol{r}\right)  _{1,\eta_{0}}\\
\mathbb{G}\left(  \boldsymbol{r}\right)  _{2,\tau_{0}} & \mathbb{G}\left(
\boldsymbol{r}\right)  _{2,\eta_{0}}%
\end{array}
\right)   &  =G_{-l}\left(  \boldsymbol{r},\boldsymbol{r}_{n}\right)  .
\tag{B12}%
\end{align}
The $S$ matrix is diagonal,%
\begin{equation}
\mathbf{S}=\text{diag}\left(  s_{\pm l}^{\left(  n\right)  }\right)
_{2N\left(  2l_{\max}+1\right)  \times2N\left(  2l_{\max}+1\right)  },
\tag{B13}%
\end{equation}
with $\mathbf{S}_{\mu_{0},\mu_{0}}=\mathbf{S}_{\lambda_{0},\lambda_{0}}%
=s_{+0}^{\left(  n\right)  }$, $\mathbf{S}_{\alpha_{0},\alpha_{0}}%
=\mathbf{S}_{\gamma_{0},\gamma_{0}}=s_{+l}^{\left(  n\right)  }$,
$\mathbf{S}_{\tau_{0},\tau_{0}}=\mathbf{S}_{\eta_{0},\eta_{0}}=s_{-l}^{\left(
n\right)  }$. Then we construct the $\mathbf{G}$ matrix, which is written as
\begin{equation}
\mathbf{G}=\left(
\begin{array}
[c]{cccc}%
0 & G\left(  1,2\right)  & \cdots & G\left(  1,N\right) \\
G\left(  2,1\right)  & 0 & \cdots & G\left(  2,N\right) \\
\vdots & \vdots & \ddots & \vdots\\
G\left(  N,1\right)  & G\left(  N,2\right)  & \cdots & 0
\end{array}
\right)  , \tag{B14}%
\end{equation}
where $G\left(  n,m\right)  $ is a $2\left(  2l_{\max}+1\right)
\times2\left(  2l_{\max}+1\right)  $ matrix, which are constructed by
$T_{l}^{-}\left[  G_{\pm l^{\prime}}\left(  n,m\right)  \right]  $.
Explicitly, one would align the $\mathbf{G}$ matrix by the following way
\begin{align}
\left(
\begin{array}
[c]{cc}%
\mathbf{G}\left(  \mu_{0},\nu_{0}\right)  & \mathbf{G}\left(  \mu_{0}%
,\beta_{0}\right) \\
\mathbf{G}\left(  \lambda_{0},\nu_{0}\right)  & \mathbf{G}\left(  \lambda
_{0},\beta_{0}\right)
\end{array}
\right)   &  =T_{0}^{-}\left[  G_{+0}\left(  n,m\right)  \right]  ,\tag{B15}\\
\left(
\begin{array}
[c]{cc}%
\mathbf{G}\left(  \mu_{0},\alpha\right)  & \mathbf{G}\left(  \mu_{0}%
,\gamma\right) \\
\mathbf{G}\left(  \lambda_{0},\alpha\right)  & \mathbf{G}\left(  \lambda
_{0},\gamma\right)
\end{array}
\right)   &  =T_{0}^{-}\left[  G_{+l^{\prime}}\left(  n,m\right)  \right]
,\tag{B16}\\
\left(
\begin{array}
[c]{cc}%
\mathbf{G}\left(  \mu_{0},\nu\right)  & \mathbf{G}\left(  \mu_{0},\beta\right)
\\
\mathbf{G}\left(  \lambda_{0},\nu\right)  & \mathbf{G}\left(  \lambda
_{0},\beta\right)
\end{array}
\right)   &  =T_{0}^{-}\left[  G_{-l^{\prime}}\left(  n,m\right)  \right]  ,
\tag{B17}%
\end{align}
for $l=0$ and $n\neq m$, and
\begin{align}
\left(
\begin{array}
[c]{cc}%
\mathbf{G}\left(  \alpha_{0},\nu_{0}\right)  & \mathbf{G}\left(  \alpha
_{0},\beta_{0}\right) \\
\mathbf{G}\left(  \gamma_{0},\nu_{0}\right)  & \mathbf{G}\left(  \gamma
_{0},\beta_{0}\right)
\end{array}
\right)   &  =T_{l}^{-}\left[  G_{+0}\left(  n,m\right)  \right]  ,\tag{B18}\\
\left(
\begin{array}
[c]{cc}%
\mathbf{G}\left(  \alpha_{0},\alpha\right)  & \mathbf{G}\left(  \alpha
_{0},\gamma\right) \\
\mathbf{G}\left(  \gamma_{0},\alpha\right)  & \mathbf{G}\left(  \gamma
_{0},\gamma\right)
\end{array}
\right)   &  =T_{l}^{-}\left[  G_{+l^{\prime}}\left(  n,m\right)  \right]
,\tag{B19}\\
\left(
\begin{array}
[c]{cc}%
\mathbf{G}\left(  \alpha_{0},\nu\right)  & \mathbf{G}\left(  \alpha_{0}%
,\beta\right) \\
\mathbf{G}\left(  \gamma_{0},\nu\right)  & \mathbf{G}\left(  \gamma_{0}%
,\beta\right)
\end{array}
\right)   &  =T_{l}^{-}\left[  G_{-l^{\prime}}\left(  n,m\right)  \right]
,\tag{B20}\\
\left(
\begin{array}
[c]{cc}%
\mathbf{G}\left(  \tau_{0},\nu_{0}\right)  & \mathbf{G}\left(  \tau_{0}%
,\beta_{0}\right) \\
\mathbf{G}\left(  \eta_{0},\nu_{0}\right)  & \mathbf{G}\left(  \eta_{0}%
,\beta_{0}\right)
\end{array}
\right)   &  =T_{l}^{+}\left[  G_{+0}\left(  n,m\right)  \right]  ,\tag{B21}\\
\left(
\begin{array}
[c]{cc}%
\mathbf{G}\left(  \tau_{0},\alpha\right)  & \mathbf{G}\left(  \tau_{0}%
,\gamma\right) \\
\mathbf{G}\left(  \eta_{0},\alpha\right)  & \mathbf{G}\left(  \eta_{0}%
,\gamma\right)
\end{array}
\right)   &  =T_{l}^{+}\left[  G_{+l^{\prime}}\left(  n,m\right)  \right]
,\tag{B22}\\
\left(
\begin{array}
[c]{cc}%
\mathbf{G}\left(  \tau_{0},\nu\right)  & \mathbf{G}\left(  \tau_{0}%
,\beta\right) \\
\mathbf{G}\left(  \eta_{0},\nu\right)  & \mathbf{G}\left(  \eta_{0}%
,\beta\right)
\end{array}
\right)   &  =T_{l}^{+}\left[  G_{-l^{\prime}}^{k}\left(  n,m\right)  \right]
, \tag{B23}%
\end{align}
for $l\neq0$ and $n\neq m$, while $\mathbf{G}\left(  n,n\right)  =0$ for
$n=m.$ $\overrightarrow{\phi}$ can be written as a $2N\left(  2l_{\max
}\mathtt{+}1\right)  \mathtt{\times}1$ vector,%
\begin{equation}
\overrightarrow{\phi}=\left(
\begin{array}
[c]{cccc}%
\phi_{1}, & \phi_{2}, & \cdots, & \phi_{N}%
\end{array}
\right)  ^{\text{T}}, \tag{B24}%
\end{equation}
where $\phi_{i}=\left[
\begin{array}
[c]{cccc}%
T_{0}^{-}\left[  \Phi\left(  \boldsymbol{r}_{i}\right)  \right]  , & T_{1}%
^{-}\left[  \Phi\left(  \boldsymbol{r}_{i}\right)  \right]  , & T_{1}%
^{+}\left[  \Phi\left(  \boldsymbol{r}_{i}\right)  \right]  , & \mathtt{\cdots
},
\end{array}
\right.  $ $\left.
\begin{array}
[c]{cc}%
T_{l_{\max}}^{-}\left[  \Phi\left(  \boldsymbol{r}_{i}\right)  \right]  , &
T_{l_{\max}}^{+}\left[  \Phi\left(  \boldsymbol{r}_{i}\right)  \right]
\end{array}
\right]  ^{\text{T}}$. Explicitly, for $l^{\prime}=0,$%
\begin{equation}
\overrightarrow{\phi}_{\nu_{0}}=P_{0}^{-}\left[  e^{i\boldsymbol{k}%
\cdot\boldsymbol{r}_{n}}/\sqrt{2}\right]  ,\overrightarrow{\phi}_{\beta_{0}%
}=iP_{1}^{-}\left[  \mp ie^{i\theta_{k}}e^{i\boldsymbol{k}\cdot\boldsymbol{r}%
_{n}}/\sqrt{2}\right]  . \tag{B25}%
\end{equation}
For $l^{\prime}>0,$%
\begin{align}
\overrightarrow{\phi}_{\alpha}  &  =P_{l^{\prime}}^{-}\left[
e^{i\boldsymbol{k}\cdot\boldsymbol{r}_{n}}/\sqrt{2}\right]  ,\overrightarrow
{\phi}_{\gamma}=iP_{l^{\prime}+1}^{-}\left[  \mp ie^{i\theta_{k}%
}e^{i\boldsymbol{k}\cdot\boldsymbol{r}_{n}}/\sqrt{2}\right]  ,\tag{B26}\\
\overrightarrow{\phi}_{\nu}  &  =P_{l^{\prime}}^{+}\left[  e^{i\boldsymbol{k}%
\cdot\boldsymbol{r}_{n}}/\sqrt{2}\right]  ,\overrightarrow{\phi}_{\beta
}=iP_{l^{\prime}+1}^{+}\left[  \pm ie^{i\theta_{k}}e^{i\boldsymbol{k}%
\cdot\boldsymbol{r}_{n}}/\sqrt{2}\right]  . \tag{B27}%
\end{align}


\end{document}